\documentclass[prl,twocolumn]{revtex4}
\usepackage[dvips]{graphics,color}
\usepackage{amsmath}
\usepackage{amsfonts}

\usepackage{amssymb}
\usepackage{amstext}
\usepackage[sort&compress]{natbib}
\newcommand{\ignore}[1]{}

\begin{document}

\title{Spin chains and channels with memory.}

\author{M. B. Plenio \& S. Virmani}
\affiliation{QOLS, Blackett Laboratory, Imperial College London,
 London SW7 2BW, UK}

\affiliation{Institute for Mathematical Sciences, Imperial College
London, 53 Exhibition Road, London SW7 2PG, UK}

\date{\today}

\begin{abstract}
In most studies of  quantum channels, it is assumed that
the errors in each use of the channel are independent.
However, recent investigations of the effect of memory 
or correlations in error have led to
speculation that non-analytic behaviour may
occur in the capacity. Motivated by these observations,
we connect the study of channels with
correlated error to the study of many-body systems.
This enables us to use many-body
theory to solve some interesting models of
correlated error. These models can display non-analyticities
analogous to quantum phase transitions.
\end{abstract}

\maketitle

An important problem in quantum information is
the determination of the {\it channel capacity} of noisy quantum
channels \cite{Nielsen C}. In a typical scenario, we wish
to send information over many uses of a noisy quantum
channel. At a cost of lowering the
information content per particle, it can
be shown that quantum error correction can essentially eliminate
all errors. This leads to quantum analogues of
the {\it channel capacity} - the optimal rate at which information
may be transferred with vanishing error in the limit of many uses. In this work we will be concerned
with the capacity for transmission of {\it quantum} information
\cite{Devetak}, which we denote ${\cal Q}({\cal{E}})$ for a channel
${\cal{E}}$.

Most works on quantum channels assume that the noise is
independent in successive transmissions. However
this is never exactly true and in many realistic systems
there can be correlations in the noise. Such channels are
termed `memory channels' and their capacity may be
significantly affected by memory effects
(see \cite{Banaszek} for a recent experiment).
Consequently, quantum memory channels have
received considerable attention recently (see e.g.
\cite{Bowen M,Macchiavello P 02,Karimipour,Kretschmann W} and
references therein). In most models that have been considered
explicitly (e.g. \cite{Macchiavello P 02}), the correlations in the
noise are modelled by a small number of {\it memory parameters}.
Initial investigations of prototype models
\cite{Macchiavello P 02,Karimipour} have suggested that the capacity may undergo sharp,
non-analytic, changes at certain
values of these parameters. These investigations have not been conclusive, however, as
with current techniques these models cannot be
analysed in the relevant setting of a very large (in fact infinite)
number of channel uses.

The main aim of this work is to show that for a variety of
interesting models this obstacle can be overcome by relating the
study of memory channels to the study of {\it many-body}
physics.\ignore{Indeed essentially all memory channels considered in
the literature can be reexpressed in the framework that we introduce.}
An advantage of the framework that we introduce is that the criticality of the
underlying many-body systems can manifest itself as non-analytic
behaviour in the channel capacity of the corresponding memory
channels, thus proving the conjectured existence of such effects
in memory channels. This suggests, as demonstrated here, that the
methods of many-body theory may be used in the study of memory
channels to obtain new results. In the first part of the paper we discuss a general
framework for the channels that we consider, discussing how a capacity formula
may be derived. We then provide an explicit bound applicable to random unitary
\footnote{Such channels have interesting
properties, see M. Gregoratti \& R.F. Werner, J. Mod. Opt.
{\bf 50}, 915 (2003)} channels in terms of a thermodynamic
quantity. In the case of dephasing channels that are random
applications of orthogonal dephasing unitaries this is exact.

{\it Mapping many-body systems to memory channels.} -- A standard
way of describing noise is to assume that each transmitted `system' particle interacts 
via a unitary $U$ with its own environment.
In order to introduce memory effects we will modify this approach by asserting that the environment
particles are initially prepared in the thermal or ground state
of an interacting many-body Hamiltonian. The spatial correlations
in the environment then lead to correlations in the noise.
We will assume that once the environment has been defined by
the parent Hamiltonian, no further dynamics occur other than
the system-environment interaction. It is important to note that
many of the noise models considered in the
literature \cite{Bowen M,Macchiavello P 02,Karimipour,Kretschmann W}
can be reexpressed in precisely this way.

For the case of a memoryless quantum channel
it has  been shown \cite{Devetak} that the quantum
capacity given by
\begin{equation}
        Q({\cal E}) = \lim_{n \rightarrow \infty}
        {I({\cal E}^{\otimes n})\over n} \label{qform}
\end{equation}
where ${\cal E}$ is the channel acting on a single transmission,
$I({\cal E}) := \sup_{\rho} \left[S({\cal E}(\rho)) -
S(I \otimes {\cal E} (|\psi \rangle \langle \psi|))\right]$
is the {\it coherent information} of the quantum channel ${\cal E}$
where $S$ denotes the von-Neumann entropy, $\rho$ is a state, and
$|\psi\rangle\langle \psi|$ is a purification of $\rho$. Finally,
${\cal E}^{\otimes n}$ represents the uncorrelated channel that
acts on $n$ inputs. For channels with correlations, the channel
on $n$ inputs ${\cal E}_n$ differs from ${\cal E}_1^{\otimes n}$,
and one has to describe the memory channel by a
sequence of channels $\{{\cal E}_n\}$, describing the action
of the channel for each number of inputs $n$. Given that eq.
(\ref{qform}) is the memoryless quantum capacity, one may anticipate that
\begin{equation}
        Q(\{{\cal E}_n\}) := \lim_{n \rightarrow \infty}
        {I({\cal E}_n)\over n} \label{qcorr}
\end{equation}
is the quantum capacity of a memory channel. While
this will certainly not be true in general, eq (\ref{qcorr})
is {\it always} an upper bound on the channel capacity
and one can derive conditions for
equality which are often satisfied:

{\em Condition for eq. (\ref{qcorr}) --}  This section may be omitted by
readers not concerned with detailed proofs. We outline
the derivation of two conditions on the many-body system
which are sufficient to demonstrate that eq.
(\ref{qcorr}) is the quantum capacity. These conditions
are satisfied by a variety
of many-body systems, including matrix product states \cite{free}, and quasi-free
bosonic systems \cite{prep}. The conditions are
derived as follows. We consider a translation invariant
chain of length $N$, split into $v=N/(l+s)$ {\it sections},
each consisting of one {\em live} block of length $l$ and
one {\em spacer} block of length $s := \delta l \ll l$.
We define two related channels: the {\it Live} channel and the {\it Product}
channel. The Live channel is
${\cal{E}}_{live} := A \rightarrow {\rm tr}_{\rm env}\{U (\rho_{L_1
L_2....L_v} \otimes A) U^\dag\}$ where $A$ is Alice's input,
$U$ is the interaction between chain and input, the
$L_1,\ldots,L_v$ label the live
blocks, and the environment is traced out.
The Product channel is defined as
${\cal{E}}_{product}:= A \rightarrow {\rm tr}_{\rm env}
\{U ((\rho^l_N)^{\otimes v} \otimes A) U^\dag\}$, where $\rho^l_N$ is the
reduced state of an individual live block.
The next three steps labelled (A),(B) and (C) construct
an analogue of the arguments made in \cite{Kretschmann W}
for forgetful channels.
{\bf (A)} {\it Showing that product classical codes are good classical codes for the Live Channel}. Given an achievable
rate $R$ for the Product channel, then $\forall\epsilon > 0:
\exists N_\epsilon$
such that for $n > N_\epsilon$ channel uses there is
a $nl$-qubit code $\{\rho_i\}_{1,\ldots \nu}$ and
decoding measurement $\{M_i\}_{1,\ldots \nu}$ for
$\nu = \lfloor 2^{nlR} \rfloor$ such that
$\forall i:\mbox{tr}\{{\cal{E}}_{product}(\rho_i) M_i\}
\geq 1 - \epsilon$.
The same procedure used for the Live channel then yields,
via the triangular inequality,
$\mbox{tr}\{{\cal{E}}_{live}(\rho_i) M_i\} \geq 1 - \epsilon -{1
\over 2} ||\rho_{L_1 L_2....L_v} - (\rho^l_N)^{\otimes v}||_1.$
This leads to our first condition on the many-body system: it turns
out that one can ensure that this error vanishes, for a choice $v=l^5$ \footnote{This
choice is for a technical reason explained in \cite{prep}.}, provided
that it can be
shown that $||\rho_{L_1 L_2....L_v} - (\rho^l_N)^{\otimes v}||_1
\leq C \, v  \, l^E \exp (-Fs)$ for positive constants $C,E,F$
(see \cite{prep}). Hence if this condition holds the classical
codes for product channels are also good codes for the Live
correlated channel.
{\bf (B)} {\it Computing achievable classical rates.} The
product channel with live block length $l$ and a total
number of spins $N=v(l+s)=l^6(1+\delta)$ has a Holevo quantity given by $\chi({\cal E}^l_N) =
\chi ({\rm tr}_{\rm env} \{U ((\rho^l_N) \otimes \bullet )
U^\dag\}),$ where the $\bullet$ acts as place holder
for the channel input and where ${\cal E}^j_X$ denotes
the effect of the full channel upon a contiguous subset of
$j \leq X$ of the input spins. As in \cite{Kretschmann W} we must now understand when this expression converges to
the regularized Holevo bound of the full memory channel as  $l\rightarrow\infty$.
Suppose that we have a spin chain of total length $l+ \Delta (l)$
where $\Delta (l)> 0$ is any function such that $\lim_{l \rightarrow \infty}
\Delta (l)/l = 0$. Using {\it subadditivity} and the {\it
Araki-Lieb} inequality we find
$\chi({\cal E}^l_{l+\Delta}) \geq
\chi({\cal E}_{l+\Delta}) - 2 \Delta\log (d)$.
Now we need
to show under which conditions this remains true if the subset
of $l$ spins is drawn from a much longer chain of length
$N=l^6(1+\delta)$. For any input $\omega$ to the live block
in question the output states will differ by at most
$||{\rm{tr}_{env}}\{U[\omega \otimes (\rho^l_{l+\Delta} -
\rho^l_{l^6(1+\delta)})]U^\dag\}||_1
\leq ||U[\omega \otimes (\rho^l_{l+\Delta} -
\rho^l_{l^6(1+\delta)})]U^\dag||_1 \leq P(l,\Delta)
:=||\rho^l_{l+\Delta} - \rho^l_{N}||_1$.
Combining this with $\chi({\cal E}^l_{l+\Delta}) \geq
\chi({\cal E}_{l+\Delta}) - 2 \Delta\log (d)$
and the Fannes inequality yields
$\lim_{l \rightarrow \infty} \chi({\cal E}^l_{l^6(1+\delta)})/l
\geq \chi_{\infty} - \lim_{l \rightarrow \infty}
2P(l,\Delta(l))\log (d)$.
This gives the second condition on the many-body system: if we can pick
$\Delta(l)$ such that $\lim_{l\rightarrow \infty}
\Delta(l)/l = 0$ and
$\lim_{l\rightarrow\infty}P(l,\Delta(l))=0$, then
the regularized Holevo quantity is
the correct classical capacity. {\bf (C)}
{\it Coherentification}. The final step is
to argue that the above arguments for classical
coding can be `coherentified' \cite{Devetak}
into a quantum code. This analysis does not
give new conditions on the many-body system
and can be conducted as
in \cite{Kretschmann W} (see \cite{prep} for details).

{\em Explicit computation of capacities --}
Even if the many-body system can be well understood,
the explicit computation of the capacity may still
be difficult, as it depends upon the interaction of
each system with its environment $U$. Judicious choice 
of $U$ will allow us to obtain analytically solvable
models. To this end we will choose $U$ to be of the
form of a controlled-PHASE gate, denoted by $U_z$,
where the environmental particles act as controls.
In this case it becomes possible to write down explicit
formulae for eq. (\ref{qcorr}) in terms of properties
of the many-body environment that share a close
relationship with thermodynamical quantities. For all
other random unitary noise the approach leads to
lower bounds, although they are not
always exact - see the concluding section for
discussion. For d-dimensional systems the controlled-PHASE
gate is defined as $U_z = \sum_{k=1}^d |k \rangle\langle k|
\otimes Z(k)$ where $Z(k):= \sum_r \exp(2\pi i kr/d) |r \rangle\langle
r|$, and the first tensor factor acts on the environment. This interaction
leads to channels that are probabilistic applications of $Z(k)$ unitaries
on the system particles, with the (correlated) probabilities determined by the diagonal
elements of the environment state.
Now eq. (\ref{qcorr}) can be written
as 
\begin{equation}
        Q(\{{\cal E}_n\}) = \log d - \lim_{n \rightarrow \infty} {S({\rm
        Diag}(\rho_{env})) \over n} \label{qkey}
\end{equation}
where ${\rm Diag}(\rho_{env})$ is the state obtained by
eliminating all off-diagonal elements of the state
of the environment in the computational basis.
Hence computing the capacity of our
channel $\{{\cal E}_n\}$ reduces to computing the
regularised diagonal entropy of the environment. Although 
this  is unlikely to be generally computable, it is amenable 
to a great deal of analysis using many-body theory.

{\it Proof sketch of equation (\ref{qkey})} The proof
utilises the {\it Choi-Jamiolkowski}
representation of the quantum channel. Given any quantum
operation ${\cal{E}}$ acting upon a $d$-level quantum
system, one may form the corresponding Choi-Jamiolkowski (CJ)
state
$J({\cal{E}}) = I \otimes {\cal{E}}(|+\rangle\langle+|)$,
where $|+\rangle = {1 \over \sqrt{d}} \sum_{i=1..d}|ii\rangle.$
The proof of equation (\ref{qkey}) follows from three steps:
(1) we argue that for the kinds of channels we consider, a
copy of $J({\cal{E}})$ allows one to physically implement
the channel exactly, (2) for any channel that can be implemented
using $J({\cal{E}})$, we argue that the quantum channel capacity
$Q({\cal{E}})$ of the channel equals $D_1(J({\cal{E}}))$, the
one-way distillable entanglement of the state $J({\cal{E}})$,
(3) we then use known results on $D_1$.

{\em Step (1) --} For simplicity we describe the argument
for channels ${\cal{E}}$ that are random applications of
Pauli rotations on a single qubit. The argument generalises
easily to Pauli channels on many qudits. Suppose that you have $J({\cal{E}})$ and you want to
implement one action of ${\cal{E}}$ upon an input state
$\rho$. This can be achieved by teleporting $\rho$ through $J({\cal{E}})$. This will leave you with a
state ${\cal{E}}(\sigma_i \rho \sigma^\dag_i )$, with the Pauli
error $\sigma_i$ depending upon the outcome of the teleportation
measurement.
As ${\cal E}$ is a random Pauli channel we can now
``undo" the error by applying the
inverse of $\sigma_i$. Hence we have $\sigma_i {\cal{E}}(\sigma_i \rho
\sigma^\dag_i ) \sigma_i = {\cal{E}}(\rho)$ and
we have implemented one action of ${\cal E}$.

{\em Step (2) --}
Our aim is to show that for channels that may be physically
implemented using $J({\cal E})$, the 1-way distillable
entanglement of $J({\cal E})$, $D_1(J({\cal E}))$, is
equivalent to $Q({\cal E})$. In \cite{distill} it was
essentially shown that  $Q({\cal E}) \geq D_1(J({\cal E}))$.
For the converse inequality consider
the specific protocol:
(a) Alice prepares many perfect EPR pairs and encodes
one half according to the code that achieves the quantum
capacity $Q({\cal{E}})$. (b) She teleports the encoded
qubits through the copies of $J({\cal{E}})$, informing
Bob of the outcome so that he can undo the effect of the
Paulis. (c) This effectively implements the channel ${\cal{E}}$ between Alice
and Bob.
(d) Bob decodes the optimal code, thereby
sharing perfect EPR pairs with Alice, at the rate determined
by $Q({\cal{E}})$. As this is a specific one-way distillation
protocol, this means that $Q({\cal{E}}) \leq D_1(J({\cal{E}}))$.
These arguments extend straightforwardly to any channel
that is a mixture of Paulis on many qudits.

{\em Step (3) --} The CJ
states of our channel are so-called {\it maximally correlated}
state, for which the distillable entanglement is known and is given by the
Hashing bound \cite{Rains}
$D_1(J({\cal E})) = S({\rm tr_B}\{J({\cal E})\}) - S(J({\cal E}))$
where $S$ is the von-Neumann entropy. Hence for such
channels ${\cal E}$ this expression is also the {\it single
copy} coherent information. In our cases we are
interested in the regularised value of this quantity,
which is given by equation (\ref{qkey}). This expression for the coherent
information has an
interesting interpretation for the dephasing interactions that we consider
- it represents the classical information lost to the environment
that is needed to correct the errors
\cite{Buscemi}.

The simplicity of eq. (\ref{qkey}) enables one to immediately
write down many models for which eq. (\ref{qcorr}) can
both be calculated, and also represents the quantum
capacity.

{\it Classical Environments --} We discuss briefly two cases.
If the environment consists of classical systems described by
a classical {\it Markov Chain} then in a large number of cases
eq. (\ref{qkey}) can be written explicitly with a simple
expression that represents the {\it entropy rate} of the
Markov chain \cite{Welsh}. Related results on Markov chain models
have been obtained using different methods in \cite{Hamada}.
In general a classical environment is represented by a diagonal
state and the second term of eq. (\ref{qkey}) is precisely
the entropy. Hence in this case the capacity becomes
\begin{equation}
        Q(\{{\cal E}_n\}) = 1 - \log_2(e)
        \left( 1 - \beta {\partial \over
        \partial \beta} \right) \lim_{n \rightarrow \infty}
        {1 \over n} \ln Z_n \label{class}
\end{equation}
where $Z_n$ is the partition function for $n$ environment spins,
$\beta = 1/k_B T$, and the $\log_2(e)$ converts from nats to
bits. Equation (\ref{class}) shows that we can now use results from 
classical statistical physics to compute the capacity - any classical spin-chain models
with sufficiently decaying correlations that can be solved
exactly will lead to memory channels that can be `solved
exactly'.

{\it Quantum Environments} -- For quantum environments eq. (\ref{qkey}) represents
the entropy that results when every environment qubit is
completely dephased. Although this quantity is not standard
in statistical physics, we expect that it may be amenable to
the techniques of many-body theory. Here we provide support
for this claim by solving analytically a class of quantum
environments inspired by recent work on
{\it Matrix Product States} (MPS) \cite{Fannes NW
92}. For MPS states the two conditions required to prove equation
(2) are satisfied except at transition points \cite{free}.

In \cite{Wolf OVC 05} it was shown that there are Hamiltonians
that exhibit quantum phase transitions and have ground states
that are MPS involving only {\it rank-1} matrices.
%
%
We will now show that for MPS involving {\it rank-1} matrices a full analytical
treatment of the quantum channel capacity of the associated
memory channel becomes possible. To this end we demonstrate
that the diagonal elements of such rank-1 MPS are given by
the probabilities of microstates in related classical Ising
chains. For simplicity we will focus on a translationally
invariant MPS for a 1D system of 2-level particles with
periodic boundary conditions. Generalisation to other
rank-1 MPS is straightforward. Such an environment state
is characterized by two matrices $Q_0$ and $Q_1$ and
is given by
$|\psi\rangle = \sum_{i_1\ldots,i_N} {\rm{tr}}\{Q_{i_1}\ldots Q_{i_N}\}
|i_1\ldots i_N\rangle.$
The {\it unnormalized} state resulting from dephasing each qubit is
\begin{equation}
\rho = \sum_{i_1\ldots,i_N} {\rm{tr}}\{\prod_{k=1}^N(Q_{i_k} \otimes Q^*_{i_k}))\} |i_1\ldots i_N\rangle\langle i_1\ldots i_N|.
\end{equation}
Relabeling the matrices $A_i=Q_i \otimes Q^*_i$, the
diagonal elements in the computational basis are of the
form {tr$\{\prod_k A_k\}$}. As the $A_i$ are both rank-1
with unique non-zero eigenvalues $a_i$, the normalised
matrices $\tilde{A}_i={A_i/a_i}$ satisfy $\tilde{A}_i^2=\tilde{A}_i$.
Using this idempotency it is easy to
show that if $|i_1\ldots i_N\rangle$
has 
$l$ occurrences of $0$ and $N-l$
occurrences of $1$, and $K$ 
boundaries between blocks of $0$s and blocks of $1$s,
then the corresponding diagonal element of $\rho$ will be
$p(l,n-l,K)=(a^{l}b^{N-l})
\mbox{tr} \{( {\tilde{A}_0\tilde{A}_1})^{K} \}/C(N)$, where $C(N)$ is a normalisation
factor. Noting that
$\tilde{A}_0\tilde{A}_1$ is also rank-1, denote its non-zero
eigenvalue by $c$. Hence the diagonal elements are $p(l,n-l,K)=a^{l}b^{N-l} c^K/C(N)$.
Hence for channels described by rank-1
MPS states, $a,b,c$ are the only relevant parameters
and we may choose {\it any} matrices with those
parameters. We choose the matrices of a {\it classical}
(i.e. diagonal) chain:
\begin{eqnarray}
  A_0 = \left(\begin{array}{cc}
     a &\sqrt{c a b} \\
     0 & 0 \end{array}\right)\;;\; A_1 = \left(\begin{array}{cccc}
     0 & 0 \\
     \sqrt{c a b} & b \end{array}\right) \label{choice}.
\end{eqnarray}
Here we have assumed that $c>0$ (which is guaranteed for
$N>5$ as otherwise the state can become non-positive). It is then easy to check that these matrices have the
correct values of $a,b,c$, as required. We choose this form because
the matrices are essentially the top row and bottom row of a {\it
transfer matrix} \cite{Sachdev} corresponding to classical Ising
chain. Roughly speaking, the parameter $c$ encodes the coupling
between adjacent antiparallel spins, and the $a,b$ encode the
coupling between adjacent parallel spins.

\begin{figure}[h]
\resizebox{7.5cm}{!}{\includegraphics{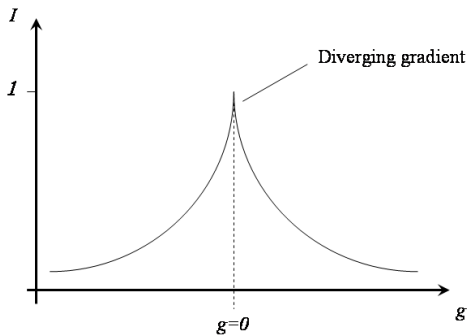}} \caption{A
sketch of the capacity for an environment that is the ground state of the Hamiltonian $\sum_i 2(g^2-1)\sigma_z^{i}\sigma_z^{i+1} - (1+g)^2 \sigma_x^{i} + (g-1)^2 \sigma_z^{i}\sigma_x^{i}\sigma_z^{i+1}$ \cite{Wolf OVC 05}. The plot's symmetry is expected as the
channel is invariant under $g \rightarrow -g$.
However, near the `phase transition' $g=0$, the gradient
diverges.} \label{wolfplot}
\end{figure}

This connection implies that the limit in eq. (\ref{qkey})
can be computed easily using well known methods \cite{Sachdev}. Fig. \ref{wolfplot}
shows the result for a Hamiltonian presented in
\cite{Wolf OVC 05}
for which the ground state is known to be a rank-1 MPS
possessing a non-standard `phase transition' at $g=0$
\cite{Fisheretal}, at which some correlation functions are
continuous but non-differentiable, while the ground state energy is
actually analytic \cite{Wolf OVC 05}.
Fig. \ref{wolfplot} shows that this is mirrored
in the non-analyticity of the channel capacity.

{\it Generalisations and Future work} -- It is important
to know whether our
approach could prove useful for other interactions. Some
generalisations are immediate. For instance, given {\it any} channels that
are probabilistic applications of unitaries, expression (\ref{qkey}) can easily be shown to be an explicit lower bound to the coherent information, and hence if the environment state has sufficiently
decaying correlations it will also be a {\it
lower} bound to channel capacity. It is likely that any channel
whose capacity can be bounded by such simple entropic
expression will benefit from similar
insights. In the long term one might speculate that there may be a
deeper explanation for these connections - not in terms of entropic
expressions appearing in both fields, but in
terms of a link between coding and
many-body physics.
\ignore{Another avenue of generalisation would be to
investigate `multidimensional' situations, in which the many-body
system has 2 or more dimensions. In such situations use of the term
`memory' is less appropriate as there is no longer only one
dimension representing time. However, the extension to
multi-dimensional systems opens new possibilities, because even for
classical spin models there are a richer variety of critical systems
in more than one dimension.}

{\it Acknowledgments} -- We are grateful to C. Macchiavello,
D. Gross, D. Kretschmann, R. Werner and M.
Hartmann for many discussions.
This work was funded by the Royal Commission for the Exhibition of
1851, EPSRC QIP-IRC, EU Integrated project QAP,
and the Royal Society.

\end{document}